\documentclass{appolb}
\usepackage{epsfig}

\begin{document}
\pagestyle{plain}
\newcount\eLiNe\eLiNe=\inputlineno\advance\eLiNe by -1
\title{STRUCTURE FUNCTION MEASUREMENTS AT LHC} 
\author{Alessandro TRICOLI on behalf of the ATLAS Collaboration
\address{University of Oxford\\ The Denys Wilkinson Building, Keble Road, Oxford, OX1 3RH, United Kingdom}
}
\maketitle

\begin{abstract}
Since the current uncertainty on the structure of the proton affects the new physics discovery potential of LHC, the ATLAS collaboration is 
investigating methods to constrain this uncertainty over the whole LHC kinematic regime. The Standard Model processes such as direct $\gamma$, $Z$, 
$W$ and inclusive jet productions are optimal candidates for this purpose. 
\end{abstract}

\section{Parton Distribution Functions ({\em PDF's}) at LHC}
PDF's, the parametrisations of the partonic content of the proton, are vital for reliable predictions for new physics signals
 and their background cross sections at the LHC.
Every cross section calculation is the convolution of the cross section at parton level
 and PDF's, $f_{i}(x, Q^2)$, where 
 $x$ is the momentum fraction of the parton involved in the hard process, $Q$ is the energy scale of the hard interaction, and $i$ represents the 
parton flavour.
Since $QCD$ does not predict the parton content of the proton,
 the PDF parameters are determined by fit to data from experimental observables in various processes, 
 using the {\em DGLAP} evolution equation. PDF's are nowadays available up to the next-to-next-to leading order ({\em NNLO}). 
Recently PDF's also provide uncertainties which take into account experimental 
errors and their correlations.
\begin{figure}[t]
  \begin{minipage}[t]{.30\textwidth} 
  	\begin{center} 
      \includegraphics[height=5.0cm, width=6.5cm]{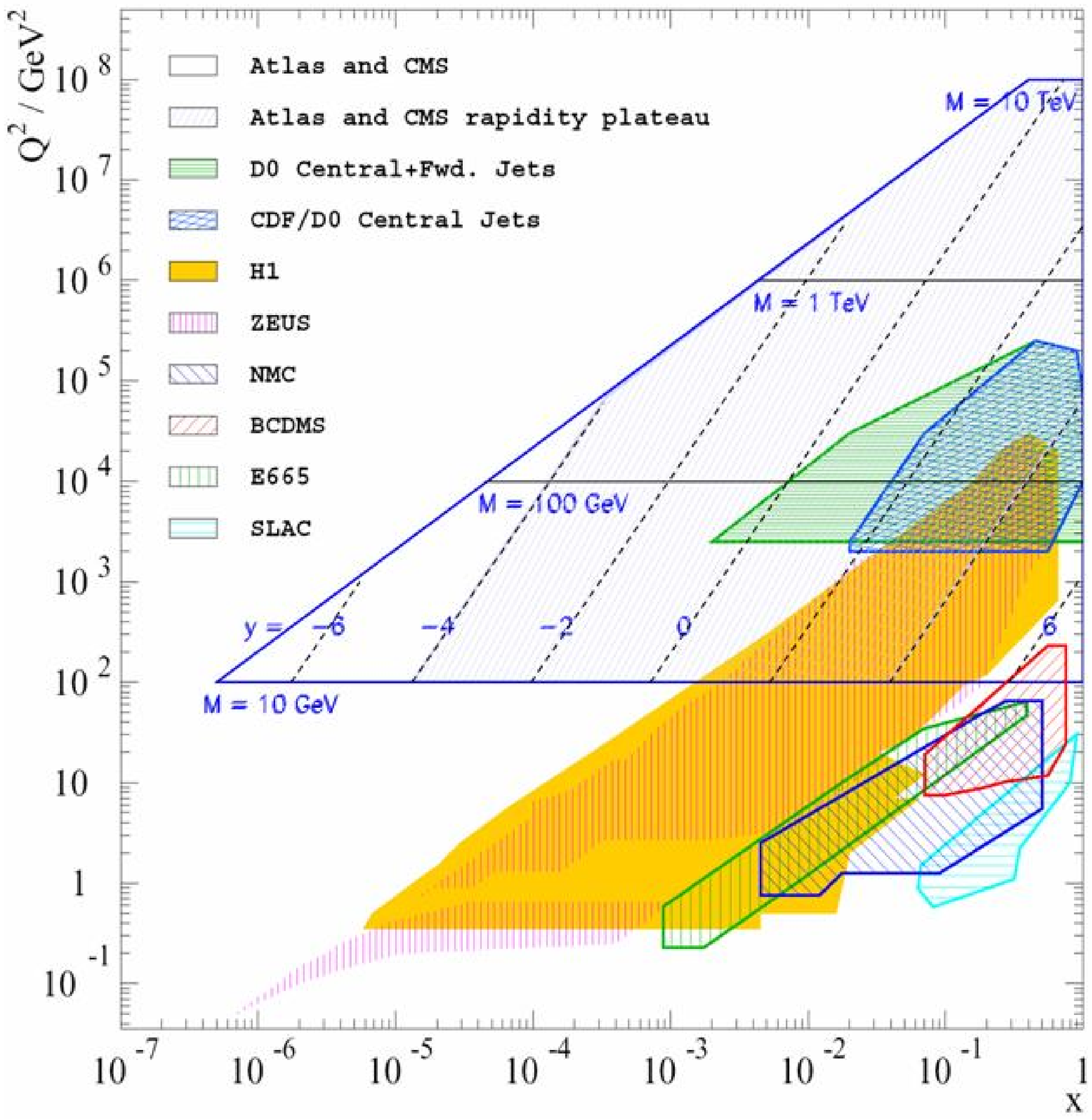}
	\end{center}
  \end{minipage}
 \hspace{26mm} 
  \begin{minipage}[t]{.30\textwidth}
    \begin{center} 
      \includegraphics[
           clip=true, viewport=10 100 530 530, height=4.2cm, width=6.5cm]
{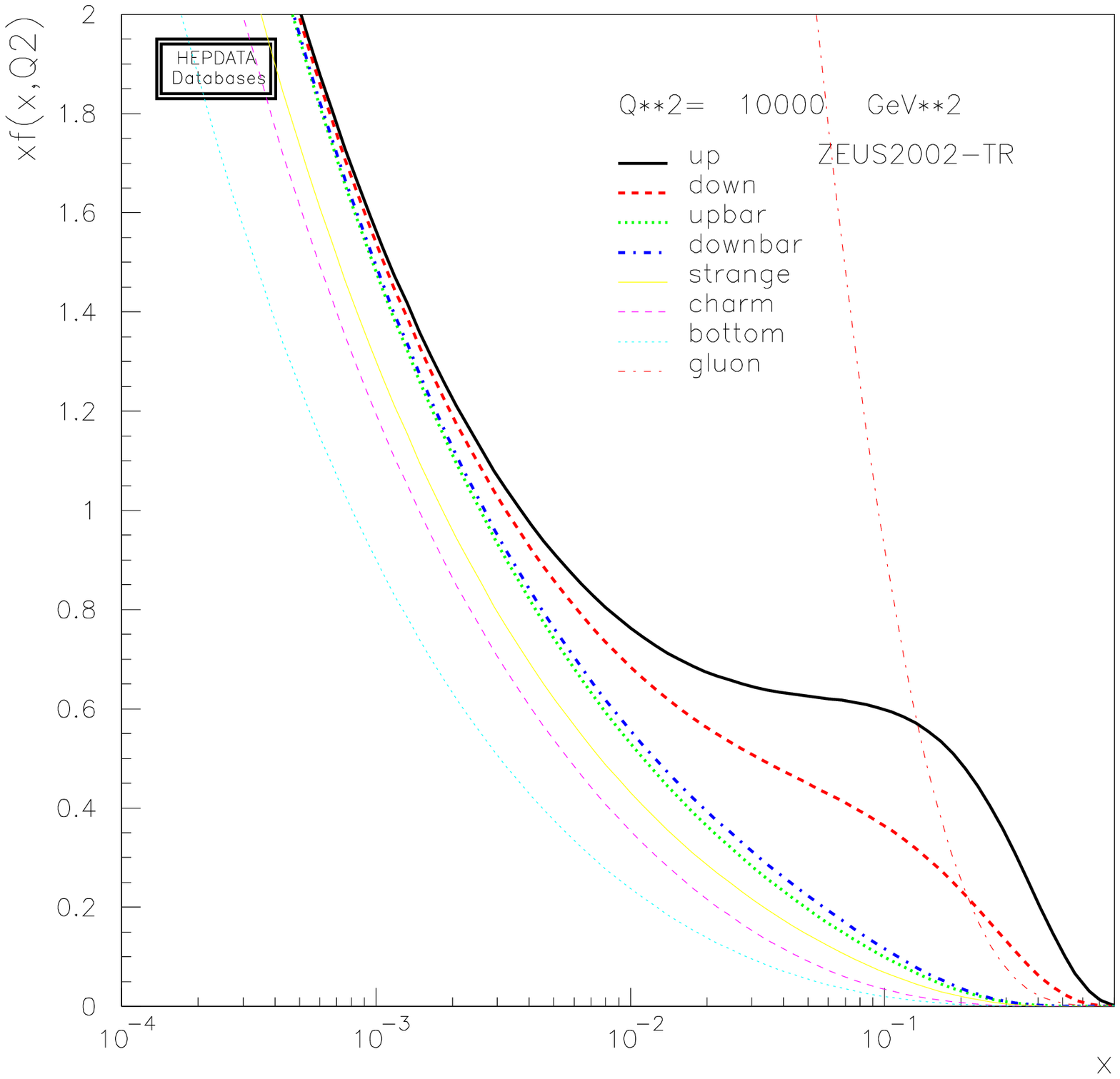}
    \end{center}
  \end{minipage}
  \caption[LHC Kinematic Regime and PDF's]{\label{fig:LHCKinRegime}}{\emph{LHC Kinematic plane (left), ZEUS-S PDF at $Q^2\approx M^2_{W,Z}$ (right).}} 
  \hfill
\end{figure}
Since the LHC kinematic region is much broader than currently explored, we will have the unique opportunity to test QCD at very $high~and~low~x$, 
where predictions are extremely important for precise measurements and new physics searches at the LHC: at the electro-weak ({\em EW}) scale  
 the theoretical predictions are dominated by {\em low-$x$ gluon} uncertainty
and at the TeV scale, where we expect new physics, they are instead dominated by the {\em high-$x$ gluon}, which is not 
well constrained by PDF fits.     

\subsection{Impact of PDF Uncertainty on ATLAS discovery potential}  
Studies~\cite{PDFError_Higgs, PDFError_ED} show that the PDF uncertainty on the Standard Model ({\em SM}) Higgs cross section prediction is $\sim 
10-15\%$ and that in some Extra Dimensions scenarios the high-$x$ gluon uncertainty
 can decrease the discovery reach in di-jet cross section measurements from $5-10~TeV$ to  
$\le 2-3~TeV$.
The high-$x$ quark uncertainty dominates the Drell-Yan cross section increasingly with the di-lepton invariant mass scale: $\sim 10\%$ at $2.5~TeV$.

\section{PDF Constraining Potential of ATLAS}
The LHC will be able to constrain the gluon PDF through the $\gamma$, $Z$, $W$ 
and inclusive jet productions.

\subsection{Direct Photon production}
At LO direct photon production will take place via the Compton scattering ($\sim 90\%$) $qg\rightarrow \gamma q$ and annihilation ($\sim 10\%$) $q 
\bar{q}\rightarrow \gamma g $ processes. The typical event topology in the ATLAS detector will be a photon and a jet, 
back-to-back in the $r-\phi$ plane. The photon and jet $p_T$ distributions are extremely sensitive to PDF differences: 
the discrepancy between different PDF sets can be of the order of $16-18\%$. This measurement could constrain the gluon distribution at mid-to-high-$x$. The challenge for the ATLAS detector is to reliably discriminate between photons and jets with a high $\gamma$ selection 
efficiency, currently estimated to be greater than $91\%$~\cite{IHollins}.

\subsection{ Inclusive jet production}
 The inclusive jet cross section is particularly sensitive to new physics,
 but experimental and theoretical errors can distort the measurements and predictions creating false signals of new physics. The experimental 
uncertainty is dominated by the jet energy scale, the main sources of theoretical uncertainties are the renormalisation, factorisation scales and the 
PDF uncertainty.  
 Recent studies~\cite{IncJets,HERA2_proj} show that the PDF uncertainty dominates as the jet $E_T$ increases and is up to $60\%$ for $E_T \sim 5~TeV$. 
This large uncertainty indicates that we can improve the high-$x$ gluon PDF with high $E_T$ jets, up to $\sim 1~TeV$, even with a $\sim 1~fb^{-1}$ 
data sample. However, this requires controlling the systematic errors such as the jet energy scale.

\subsection{$W$ and $Z$ rapidity distributions}
\begin{figure}[t]
  \begin{minipage}[t]{.30\textwidth} 
  	\begin{center} 
	  \includegraphics[clip=true,
	    viewport=10 10 500 500, height=6.7cm, width=7.8cm] {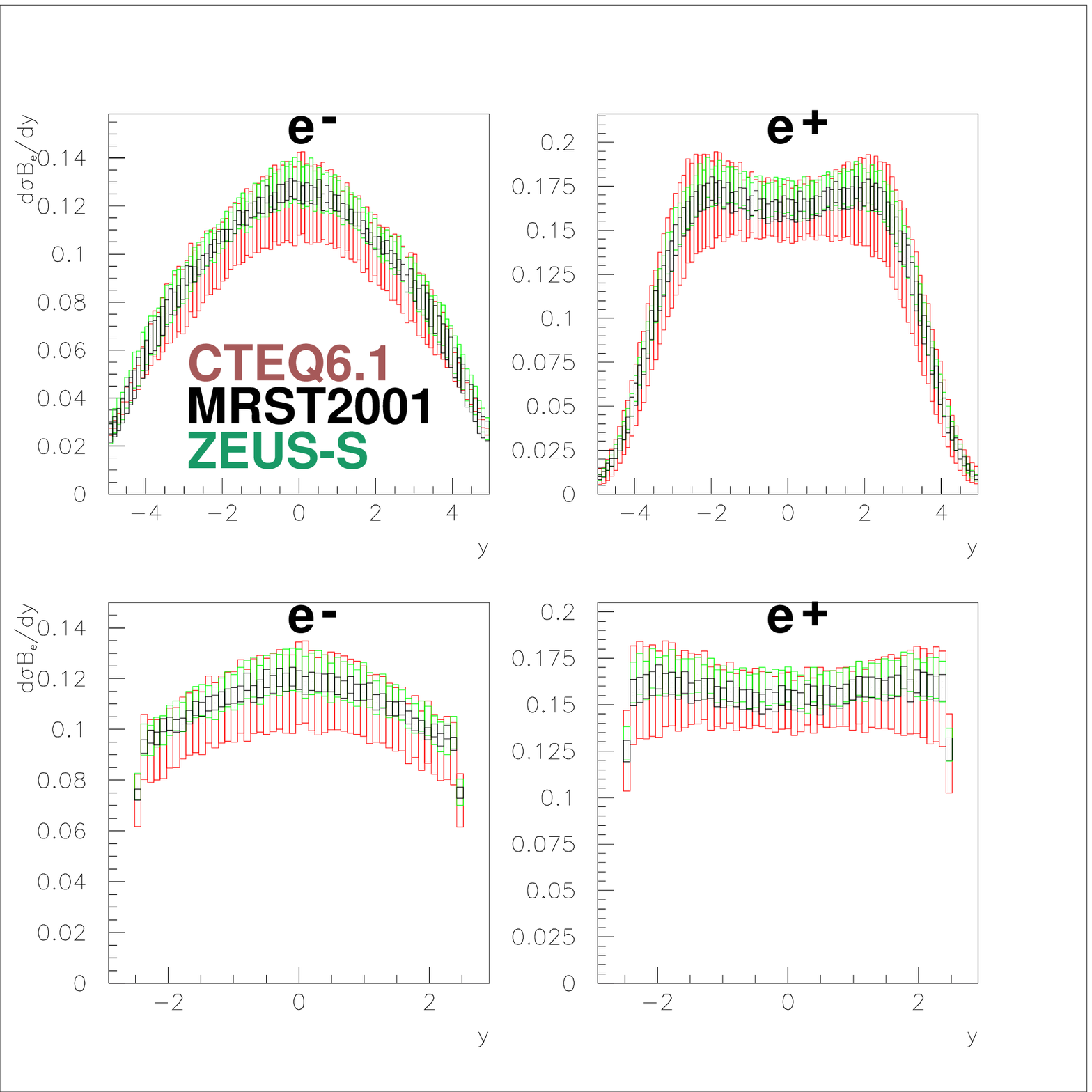} 
	\end{center}
  \end{minipage}
 \hspace{42mm} 
  \begin{minipage}[t]{.30\textwidth}
    \begin{center}  
      \includegraphics[clip=true, 
           viewport=10 10 500 500, height=6.7cm, width=4.0cm]{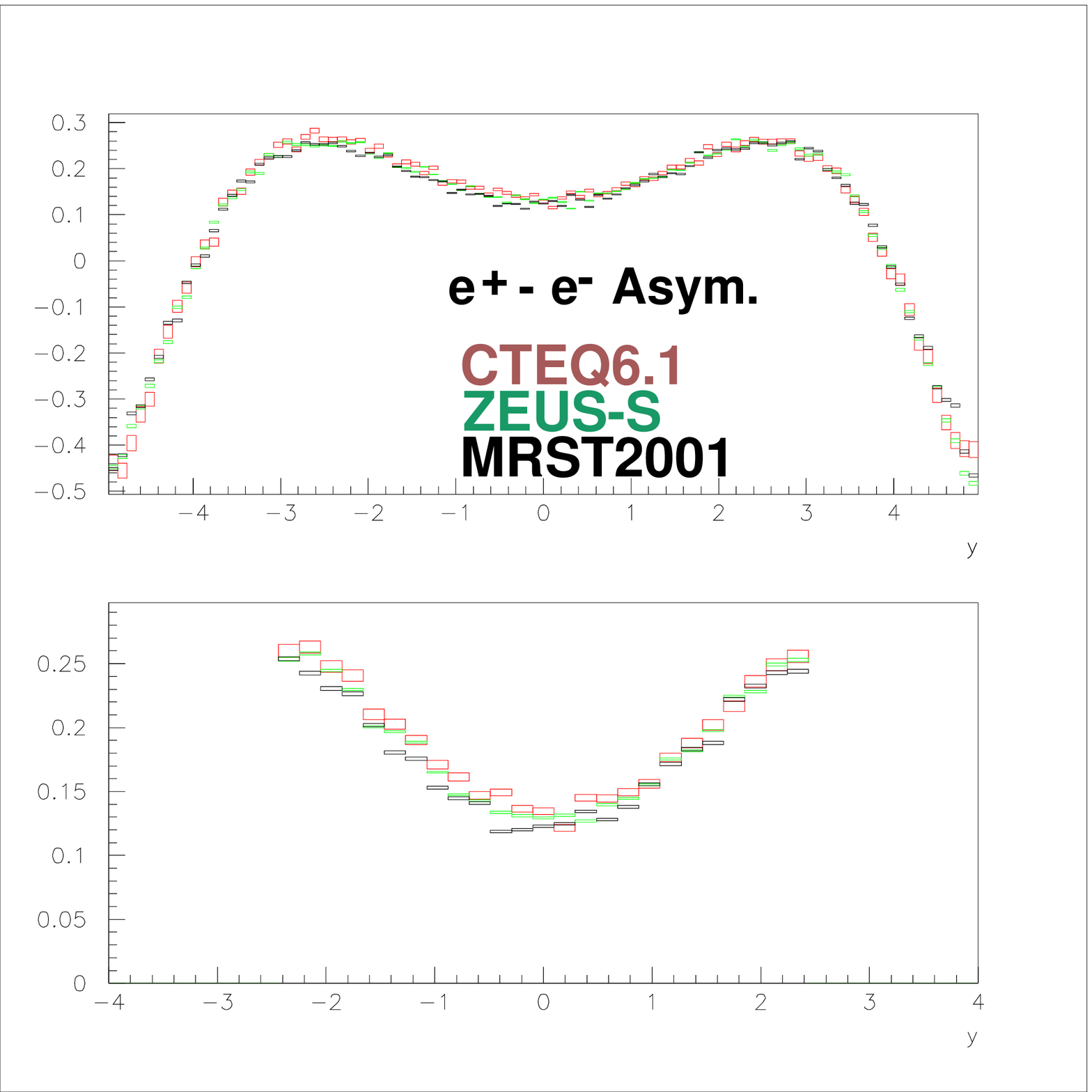}
    \end{center}
  \end{minipage}
  \caption[$e^+ e^-$ rapidity distributions at generator and detector level and Asymmetry distribution]{\label{fig:eRapDistrAndAsym}}{\emph{
Lepton rapidity spectra from $W$ decay with PDF uncertainties: at HERWIG MC level (top), after the event selection in the ATLAS detector (bottom).}}
  \hfill
\end{figure}

The uncertainties on $Z$ and $W$ boson production cross sections are dominated by the PDF uncertainty: the $Z$ and $W$ rapidity distributions are 
theoretically known to NNLO~\cite{NNLO_WZ_y},
with residual scale dependence $<1\%$, whereas
the total PDF uncertainty at $|y|<2.5$, corresponding to $10^{-4}<x<10^{-1}$, is $\sim 8\%$~\cite{ourpaper_HERAtoLHC} (see 
fig.~\ref{fig:eRapDistrAndAsym}). The $Z$ and $W$ are very clean signals at the LHC: the background contamination on $W$ events can be reduced to 
$\sim 1\%$. 
The PDF's precision may be strikingly improved by ATLAS if the detector systematic uncertainties can be controlled to the $\sim 4\%$ level.
To estimate the ATLAS contribution to a global PDF fit we included ATLAS {\em pseudo-data} into the ZEUS-S PDF fit: as a result
the error on the parameter $\lambda$, which controls the low-$x$ gluon, $xg(x)\approx x^{-\lambda}$, is reduced by $35\%$.\\
The $W$ asymmetry can be considered a SM benchmark for LHC, since it cancels out detector and some PDF uncertainties down to $4\%$.        

\subsection{ $Z + b$-$jet$ production}
The $bg \rightarrow Zb$ process is sensitive to the $b$-quark content of the proton and is also a background to the Higgs boson 
search~\cite{Campbell_1,Campbell_2}. The accurate determination of the $b$-PDF is important for the precise measurement of the $Z$ boson production 
cross section: 
in order to measure $\sigma_Z$ to about $1\%$, a $b$-PDF precision of $\sim 20\%$ is required.
For ATLAS, the prediction of the $Z+b$ cross-section using different PDF sets is $\pm 5-10\%$.
In the $Z\rightarrow \mu^+ \mu^-$ channel we can obtain a clear signal over the background, with a $Z+b$ event selection efficiency of $\sim 15\%$ and 
sample purity $53\pm 10\%$~\cite{VerducciEtAl}. 

\section{Conclusions and Outlook}
Precision PDF's are crucial at LHC as they can compromise the potential for new physics discovery. 
The SM processes like {\em direct photon, Z, W and inclusive jet productions} are optimal to constrain PDF's at LHC, in particular the {\em gluon} 
parameters.
These measurements are not limited by statistics but by systematics and in order to constrain the gluon parameters to unprecedented precision  we must 
keep the experimental systematics down to $4\%$.  
 There are also indications that before the LHC start up in 2007 the HERA-II measurements will significantly improve our knowledge on PDF's, 
especially in the high-$x$ region which affects the high energy scale of LHC, where we expect new physics~\cite{HERA2_proj}.

\section{Acknowledgements}
I am grateful to A. Cooper-Sarkar and the Oxford group for their help and support and to the ATLAS Speakers Committee for the invitation.


\end{document}